\documentclass[aps,preprint,superscriptaddress]{revtex4}
\usepackage{epsfig}
\usepackage{mathrsfs}

\usepackage{amsfonts}
\usepackage{color}
\usepackage{amssymb}

\usepackage{siunitx}
\usepackage{amsfonts}
\usepackage{amsmath}
\usepackage{nicefrac}
\usepackage{physics} % für Differntiale und ähnliches
\usepackage{slashed}
\allowdisplaybreaks

\usepackage{graphicx}
\usepackage{subdepth} % macht eine schönes hoch Zeichen

\begin{document}

\newcommand{\NN}{\mathbb{N}}
\newcommand{\ZZ}{\mathbb{Z}}
\newcommand{\QQ}{\mathbb{Q}}
\newcommand{\RR}{\mathbb{R}}
\newcommand{\CC}{\mathbb{C}}

\newcommand{\ii}{\operatorname{i}}
\renewcommand{\vec}[1]{\ensuremath{\boldsymbol{#1}}}

% Command for the sum of the finite volume correction
\newcommand{\SumFinCorr}[1][n]{{\ensuremath{\sum_{\vec{#1}\in\ZZ^3}^{\vec{#1}\neq\vec{0}}}}}
% four and three momentum integral
\newcommand{\IntFourDPi}[1]{\ensuremath{\int\frac{\dd[4]{#1}}{(2\pi)^4}\,}}
\newcommand{\IntThreeDPi}[1]{\ensuremath{\int\frac{\dd[3]{#1}}{(2\pi)^3}\,}}

 %%%
 % Neue Befehle für die Impulse 
 %%%
 \newcommand{\El}[1][]{l_0^{#1}}
 \newcommand{\kl}[1][]{\vec{l}^{\,#1}}
\newcommand{\klI}[1]{\vec{l}_{\,#1}}

 \newcommand{\ppf}[1][]{\ensuremath{\bar{p}^{#1}}}
 \newcommand{\kf}[1][]{\ensuremath{\vec{\bar{p}}^{\,#1}}}
 \newcommand{\kfI}[1]{\ensuremath{\vec{\bar{p}}_{\,#1}}}
 \newcommand{\Ef}[1][]{\ensuremath{\bar{p}_0^{#1}}}
 
 \newcommand{\ppi}[1][]{\ensuremath{p^{#1}}}
 \newcommand{\ki}[1][]{\ensuremath{\vec{p}^{\,#1}}}
 \newcommand{\kiI}[1]{\ensuremath{\vec{p}_{#1}}}
 \newcommand{\Ei}[1][]{\ensuremath{p_0^{#1}}}

%Befehle mit q
\newcommand{\Nq}{\ensuremath{q_0}}
\newcommand{\NqP}[1]{\ensuremath{q_0^{#1}}}
\newcommand{\Vq}{\ensuremath{\vec{q}}}
\newcommand{\VqP}[1]{\ensuremath{\vec{q}^{\,#1}}}
 
 \newcommand{\tM}{t_0^-}
 \newcommand{\tP}{t_0^+}

 \newcommand{\pfi}[1][]{\ensuremath{q^{#1}}}
 \newcommand{\Efi}[1][]{\ensuremath{q_0^{#1}}}
 \newcommand{\Qfi}[1][]{\ensuremath{Q_0^{#1}}}
 \newcommand{\kfi}[1][]{\ensuremath{\vec{q}^{\,#1}}}
 
 \newcommand{\kfiI}[1]{\ensuremath{\vec{q}_{#1}}}
 
 \newcommand{\Kfi}[1][]{\ensuremath{\vec{Q}^{\,#1}}}

\newcommand{\Vx}[1][n]{\ensuremath{\vec{x}_{#1}}}
\newcommand{\VxN}[1][n]{\ensuremath{\lvert\vec{x}_{#1} \rvert}}
\newcommand{\VxI}[1]{\ensuremath{\vec{x}_{n}^{(#1)}}}
\newcommand{\VxP}[1]{\ensuremath{\vec{x}_{n}^{#1}}}

%%%
%command for AA integrals
\newcommand{\FVEAA}[4]{\ensuremath{\Delta_{V}A{}^{#1,#2}_{#3,#4}}}
\newcommand{\FVEAAL}[3]{\ensuremath{\Delta_{V}A{}^{3,#1}_{#2,#3}}}
\newcommand{\FVEAAQ}[3]{\ensuremath{\Delta_{V}A{}^{2,#1}_{#2,#3}}}
\newcommand{\FVEAAR}[3]{\ensuremath{\Delta_{V}A{}^{1,#1}_{#2,#3}}}
%%
% command for BB integrals
\newcommand{\FVEBB}[4]{\ensuremath{\Delta_{V}B{}^{#1,#2}_{#3,#4}}}
\newcommand{\FVEBBLQ}[3]{\ensuremath{\Delta_{V}B{}^{1,#1}_{#2,#3}}}
\newcommand{\FVEBBLR}[3]{\ensuremath{\Delta_{V}B{}^{2,#1}_{#2,#3}}}
\newcommand{\FVEBBQR}[3]{\ensuremath{\Delta_{V}B{}^{3,#1}_{#2,#3}}}
%%
% command for the CC integral 
\newcommand{\FVECC}[3]{\ensuremath{\Delta_{V}C{}^{#1}_{#2,#3}}}
%%
%command for the X_{1-3,a}_{k,i} integrals
\newcommand{\IntX}[4]{\ensuremath{X{}^{#1,#2}_{#3,#4}}}
\newcommand{\IntXQ}[3]{\ensuremath{X{}^{2,#1}_{#2,#3}}}
\newcommand{\IntXR}[3]{\ensuremath{X{}^{1,#1}_{#2,#3}}}
\newcommand{\IntXL}[3]{\ensuremath{X{}^{3,#1}_{#2,#3}}}
%%%
%command for the Y^{1-3,a}_{f,i} integrals
\newcommand{\IntYQR}[3]{\ensuremath{Y^{3,#1}_{#2,#3}}}
\newcommand{\IntYLQ}[3]{\ensuremath{Y^{1,#1}_{#2,#3}}}
\newcommand{\IntYLR}[3]{\ensuremath{Y^{2,#1}_{#2,#3}}}

%commands basis integrals
\newcommand{\FVEIRa}[2][R]{\ensuremath{\Delta_V I_{#1}^{(#2)}}}
\newcommand{\FVEIQa}[2][Q]{\ensuremath{\Delta_V I_{#1}^{(#2)}}}
\newcommand{\FVEILa}[2][L]{\ensuremath{\Delta_V I_{#1}^{(#2)}}}

\newcommand{\FVEIQafi}[4][Q]{\ensuremath{\Delta_V I^{(#2)}_{#1,(#3 \, #4)}}}
\newcommand{\FVEILafi}[4][L]{\ensuremath{\Delta_V I^{(#2)}_{#1,(#3 \, #4)}}}
\newcommand{\FVEIRafi}[4][R]{\ensuremath{\Delta_V I^{(#2)}_{#1,(#3 \, #4)}}}

\newcommand{\FVEILQa}[2][LQ]{\ensuremath{\Delta_V I_{#1}^{(#2)}}}
\newcommand{\FVEILRa}[2][LR]{\ensuremath{\Delta_V I_{#1}^{(#2)}}}
\newcommand{\FVEIQRa}[2][QR]{\ensuremath{\Delta_V I_{#1}^{(#2)}}}

\newcommand{\IntVLQa}[2][LQ]{\ensuremath{V_{#1}^{(#2)}}}
\newcommand{\IntVLRa}[2][LR]{\ensuremath{V_{#1}^{(#2)}}}
\newcommand{\IntVLRaReg}[3][LR]{\Bigg[\frac{\IntVLRa[#1]{#2}}{\Efi[#3]}\Bigg]_{\text{reg}}}
\newcommand{\IntELRa}[2][LR]{\ensuremath{E_{#1}^{(#2)}}}
\newcommand{\IntVQRa}[2][QR]{\ensuremath{V_{#1}^{(#2)}}}

\newcommand{\FVEILQRa}[2][LQR]{\ensuremath{\Delta_V I_{#1}^{(#2)}}}

% commands for comments
\newcommand{\FH}[1]{\textcolor{red}{[FH: #1]}}
\newcommand{\TI}[1]{\textcolor{red}{[TI: #1]}}
\newcommand{\ML}[1]{\textcolor{red}{[ML: #1]}}

%commans for infite integrals
\newcommand{\InfIQa}[2][Q]{\ensuremath{I_{#1}^{(#2)}}}
\newcommand{\InfILa}[2][L]{\ensuremath{I_{#1}^{(#2)}}}
\newcommand{\InfIRa}[2][R]{\ensuremath{I_{#1}^{(#2)}}}
\newcommand{\InfIRTa}[2][R]{\ensuremath{\tilde{I}_{#1}^{(#2)}}}

\newcommand{\InfIQ}[1][Q]{\ensuremath{I_{#1}}}
\newcommand{\InfIL}[1][L]{\ensuremath{I_{#1}}}
\newcommand{\InfIR}[1][R]{\ensuremath{I_{#1}}}

\newcommand{\InfILQR}[1][LQR]{\ensuremath{I_{#1}}}

\newcommand{\genNpiR}[4][LQR]{\ensuremath{ J_{#1}^{(#2,#3,#4)}}} %%% Änderung der Indexnotation kfi ->fki \newcommand{\FVECC}[3]{\ensuremath{\Delta_V C^{#1}_{#2,#3}}}

\newcommand{\genpiN}[4][$pi$N]{\ensuremath{ J_{#1}^{(#2,#3,#4)}}} %%% Änderung der Indexnotation kfi ->fki \newcommand{\FVEBBQR}[3]{\ensuremath{\Delta_V B^{3,#1}_{#2,#3}}}
\newcommand{\genNpi}[4][Npi]{\ensuremath{ J_{#1}^{(#2,#3,#4)}}} %%% Änderung der Indexnotation kfi ->fki \newcommand{\FVEBBLQ}[3]{\ensuremath{\Delta_V B^{1,#1}_{#2,#3}}}
\newcommand{\genLR}[4][LR]{\ensuremath{ J_{#1}^{(#2,#3,#4)}}} %%% Änderung der Indexnotation kfi ->fki \newcommand{\FVEBBLR}[3]{\ensuremath{\Delta_V B^{2,#1}_{#2,#3}}}

\newcommand{\genL}[4][L]{\ensuremath{ J_{#1}^{(#2,#3,#4)}}} %%% Änderung der Indexnotation kfi ->fki \newcommand{\FVEAAL}[3]{\ensuremath{\Delta_V A^{3,#1}_{#2,#3}}}
\newcommand{\genQ}[4][Q]{\ensuremath{ J_{#1}^{(#2,#3,#4)}}} %%% Änderung der Indexnotation kfi ->fki \newcommand{\FVEAAQ}[3]{\ensuremath{\Delta_V A^{2,#1}_{#2,#3}}}
\newcommand{\genR}[4][R]{\ensuremath{ J_{#1}^{(#2,#3,#4)}}} %%% Änderung der Indexnotation kfi ->fki \newcommand{\FVEAAR}[3]{\ensuremath{\Delta_V A^{1,#1}_{#2,#3}}}

\title{How much strangeness is needed for \\the axial-vector form factor of the nucleon?}

\author{Felix Hermsen}
\affiliation{GSI Helmholtzzentrum f\"ur Schwerionenforschung GmbH, \\Planckstra\ss e 1, 64291 Darmstadt, Germany}
\affiliation{Van Swinderen Institute for Particle Physics and Gravity, University of Groningen, 9747 Groningen, AG, The Netherlands}
\author{Tobias Isken}
\affiliation{GSI Helmholtzzentrum f\"ur Schwerionenforschung GmbH, \\Planckstra\ss e 1, 64291 Darmstadt, Germany}
\affiliation{Helmholtz Forschungsakademie Hessen f\"ur FAIR (HFHF), Campus Darmstadt, Germany}
\author{Matthias F.M. Lutz}
\affiliation{GSI Helmholtzzentrum f\"ur Schwerionenforschung GmbH, \\Planckstra\ss e 1, 64291 Darmstadt, Germany}
\author{David Thoma}
\affiliation{GSI Helmholtzzentrum f\"ur Schwerionenforschung GmbH, \\Planckstra\ss e 1, 64291 Darmstadt, Germany}
\affiliation{Technische Universit\"at Darmstadt, D-64289 Darmstadt, Germany}

\begin{abstract}

We consider the axial-vector together with its induced pseudo-scalar form factor of the nucleon as computed  from the chiral Lagrangian with nucleon and isobar degrees of freedom. The form factors are evaluated at the one-loop level, where particular emphasis is put on the use of on-shell masses in the loop expressions. Our results are presented in terms of a novel set of basis functions that generalize the Passarino--Veltman scheme to the case where power-counting violating structures are to be subtracted. The particularly  important role of the isobar degrees of freedom is emphasized. We obtain a significant and simultaneous fit to the available Lattice  QCD results based on flavour SU(2) ensembles for the baryon masses and form factors up to pion masses of about 500 MeV. Our fits includes sizeable finite volume effects that are implied by using in-box values for the hadron masses entering our one-loop expressions. We conclude that from flavour SU(2) ensembles it appears not possible to predict the empirical formfactor at the desired precision. Effects from strange quarks  are expected to remedy the situation.
\end{abstract}

\maketitle

\section{Introduction}	

The flavor SU(2) chiral Lagrangian properly formulated with nucleon and isobar degrees of freedom plays an important role in the understanding of Lattice QCD results on the form factors of the nucleon \cite{Jenkins:1990jv,Jenkins:1991es,Fuchs:2003vw,Procura:2006gq,Ledwig:2011cx,Yao:2017fym,Lutz:2020dfi}.
Such a framework is designed to be applied to Lattice QCD ensembles at fixed physical heavy-quark masses, but unphysical values of the masses of the up and down quarks \cite{Khan:2006de,Beane:2004rf,Bali:2014nma,Capitani:2017qpc,Alexandrou:2017hac,Bali:2018qus}.

The use of the flavor SU(2) chiral Lagrangian with nucleon and isobar degrees of freedom has a long history \cite{Jenkins:1990jv,Jenkins:1991es,Fuchs:2003vw,Procura:2006gq,Ledwig:2011cx,Yao:2017fym,Lutz:2020dfi}. Still, there is some controversy as to what is the most effective framework to tackle such systems.  Different assumptions on how to include and consider the isobar field are possible \cite{Procura:2006gq,Yao:2016vbz,Lutz:2020dfi,Alvarado:2021ibw}. Here Lattice QCD simulations are expected to help identifying the optimal framework.   

At this stage most useful are somewhat older Lattice QCD data on flavor SU(2) ensembles  \cite{Capitani:2017qpc,Alexandrou:2017hac,Bali:2018qus} as analyzed already by our group in \cite{Lutz:2020dfi}. This is so since here the convergence properties of the chiral approach are not affected by possible intricate strange quark mass effects. One should keep in mind, however, that one should not  expect to obtain from such Lattice QCD data an accurate reproduction of the empirical form factor at physical up, down and strange quark masses. It would be quite a surprise if the strange quarks do not play any role.  
An evaluation of the axial-vector form factor showed that the impact of the pion-isobar loop effects is significant. Therewith a large sensitivity on the used framework was illustrated. The success of the study \cite{Lutz:2020dfi}
rests on the novel feature of insisting on the use of on-shell hadron masses inside the loop expressions. Application of more conventional approaches to such SU(2) flavor lattice data have not been documented in the literature so far.
% This may be taken as hint that they are not very effective. 
Since our approach is not main-stream yet, it is nevertheless useful to scrutinize it against further quantities measurable on Lattice QCD ensembles.  
We note however, that there are some works \cite{LHPC:2010jcs,Djukanovic:2024krw} based on flavor SU(3) ensembles that 
applied more conventional chiral extrapolation techniques like proposed in \cite{Procura:2003ig}. In this case, however, an application of flavor SU(2) chiral extrapolation formulae may be questioned since the LEC will have a dependence on the not-so-well known variation of the chiral strange quark masses of such ensembles.

In the present work we will consider the induced pseudo-scalar form factor of the nucleon. It offers a unique further test-bed of our approach as it is largely determined by the set of low-energy constants (LEC) that enter the axial-vector form factor already. 
For the first time we derive results in terms of a novel generalization of the Passarino--Veltman scheme that permits a systematic subtraction of power-counting structures without generating kinematical singularities nor acausal structures 
\cite{Isken:2023xfo}. A chiral expansion is implied by using approximated coefficient functions in front of the generalized basis loop functions that are evaluated in terms of on-shell hadron masses rather than bare masses.

\section{The induced pseudo-scalar form factor of the nucleon}

The axial current in a nucleon state is parameterized by an axial-vector, $G_A(q^2)$, and an induced pseudo-scalar, $G_P(q^2)$, form factor factor 
\begin{eqnarray}
&& \bra{N(\bar{p})}A_{i}^{\mu}(0)\ket{N(p)}=
\bar{u}_{N}(\bar{p})\,F_A^\mu (\bar p, p)\,\frac{\tau_i}{2}\,u_N(p) \,,
\nonumber\\
&& F_A^\mu(\bar p, p)\big|_{\rm on\, shell} =\Big(\gamma^{\mu}\,G_A(q^2)+
\frac{q^{\mu}}{2\,M_N} \, G_P(q^2)\Big) \,\gamma_5\, ,
\nonumber\\
&& G_P(q^2) = 
\frac{8\sqrt{2} \, M_N^4}{(d-2)\,(4\,M_N^2-t)\,t} \, 
\Tr\bigg[\gamma_{\mu} \, \gamma_5 \, 
\frac{\bar{\slashed{p}}+M_N}{2 \, M_N} \, F_A^{\mu}(\bar p,p) \, \frac{\slashed{p}+M_N}{2 \, M_N}\bigg]
\nonumber\\
&& \quad \;\,\, -\,\frac{8\sqrt{2}\,M_N^4\big(4\,M_N^2\,(d-1) - t\,(d-2)\big)}
{(d-2)(4M_N^2-t)t^2}
\Tr\bigg[\frac{q_\mu}{2\,M_N} \, \gamma_5\;
\frac{\bar{\slashed{p}}+M_N}{2 \, M_N} \,  F_A^{\mu}(\bar p,p)\, \frac{\slashed{p}+M_N}{2 \, M_N}\bigg] \, ,
\label{Axialcurrent}
\end{eqnarray}
with  $q^2=(\bar p-p)^2$ and $\bar p^2=p^2=M_N^2$ (see e.g.   \cite{GASSER1988779}). Using exact isospin symmetry the form factors 
are introduced in terms of the conventional Pauli matrices $\tau_i$. 
They can be conveniently expressed as a Dirac trace over the amplitude $F_A^\mu(\bar p, p)$ in space-time dimensions $d$, where we recall the form relevant for our current work \cite{Lutz:2020dfi}. 

The chiral Lagrangian with nucleon and isobar degrees of freedom was developed in a series of works \cite{Bernard:1993bq,Bernard:1998gv,Fearing:1997dp,Schindler:2006it, Fuchs:2003qc, Chen:2012nx, Ando:2006xy,Hemmert:2003cb,Procura:2006gq,Yao:2017fym, Ellis:1997kc,Lutz:2020dfi}. Initially the  heavy-baryon chiral perturbation theory \cite{Bernard:1993bq,Bernard:1998gv,Fearing:1997dp} was applied. The relativistic form of the chiral Lagrangian was used in \cite{Schindler:2006it,Fuchs:2003qc,Chen:2012nx,Ando:2006xy}. Less well explored is the role of the isobar degrees of freedom  \cite{Hemmert:2003cb,Procura:2006gq,Yao:2017fym,Ellis:1997kc,Lutz:2020dfi}. We will use the conventions of \cite{Lutz:2020dfi,Sauerwein:2021jxb,Isken:2023xfo}, in which a renormalization scheme based on a generalized   Passarino--Veltman reduction scheme \cite{Passarino:1978jh} was propagated. Altogether, at the one-loop level we find
\begin{eqnarray}
&& F_A^\mu(\bar p,p) = \Big(  \frac{1}{\sqrt{2}} \,g_A\, \gamma^\nu \,\gamma_5 \Big)
\Big(Z_N\, g_{\nu}^{\;\mu} - \frac{q_\nu \,q^\mu}{q^2 -m_\pi^2}\,\Big[ Z_N+f_\pi/f +Z_\pi - 2\Big] \Big)
\nonumber\\
&& \quad\;\,+ \, \frac{1}{\sqrt{2}}\,\Big(
g_R \, q^2 +4 \, g^+_\chi \,( 2\,B_0\,m ) \Big) \,\gamma^\mu \, \gamma_5 - \frac{2\,M}{\sqrt{2}}\,\Big( 
4\,g_\chi ^++ g_\chi^- \Big)\,\gamma_5\,\frac{q^\mu \,(2\,B_0\,m)}{q^2-m_\pi^2} - \frac{1}{\sqrt{2}}\,g_R\,\slashed{q}\,\gamma_5\,q^\mu
\nonumber\\
&& \quad\;\, +\,\frac{g_A}{f^2}\,  \Big\{J^\mu_{\pi}(\bar p, p)  +  J_{\pi N}^\mu(\bar p, p) + J_{ N \pi }^\mu(\bar p, p) \Big\}
+ \frac{g_A^3}{4\,f^2}\,J^\mu_{N\pi N}(\bar p, p)  
+ \frac{5\,h_A\,f_S^2}{9\,f^2}\,J^\mu_{\Delta \pi \Delta}(\bar p, p) \,,
  \nonumber\\
&&   \quad \;\,+\, \frac{f_S}{3\,f^2}\,  \Big\{ J_{\pi \Delta}^\mu (\bar p, p) + J_{ \Delta \pi }^\mu(\bar p, p) \Big\} + 
 \frac{2\,g_A\,f_S^2}{3\,f^2}\,\Big\{J^\mu_{N\pi \Delta }(\bar p, p) + J^\mu_{\Delta \pi N }(\bar p, p) \Big\} +{\mathcal O} \left( Q^4\right)\,,
 \label{res-FA}
 \end{eqnarray}
where we encounter some low-energy constants (LEC) and loop integrals $J^\mu_{\cdots}(\bar p, p)$. 
Contributions from two-loop diagrams are relevant at chiral order $Q^4$. 
Unlike in our previous works on $G_A(q^2)$ our focus here is on the induced pseudo-scalar form factor $G_P(q^2)$. In turn additional contributions are required that were not considered by us before. 
The wave-function factor, not only for the nucleon, $Z_N$, but also for pion  $Z_\pi$ together with the LEC $l_4$ is encountered. We use the one-loop expression as given by \cite{Schindler:2006it} in equations (33-35) with
\begin{eqnarray}
&& Z_\pi = 1 + \frac{2}{3\,f^2}\,\bar I_\pi -2\, \frac{m_\pi^2}{f^2}\,l_4  \,,\qquad \qquad \qquad 
 f_\pi = f\, - \frac{1}{f} \,\bar I_\pi +\frac{m_\pi^2}{f}\, l_4  \,,
 \nonumber\\
&& \bar I_\pi = \frac{m_\pi^2}{(4\,\pi)^2}\,\log \frac{m_\pi^2}{\mu^2} \,, \qquad \qquad \qquad 
m_\pi^2= 2\,B_0\,m + \frac{m_\pi^2}{2\,f^2} \,\bar I_\pi
 -\frac{m_\pi^4}{2\,f^2}\,l_3\,,
 \end{eqnarray}
in terms of the renormalization scale $\mu$ of dimensional regularization. Explicit expressions for the form of the nucleon wave function factor $Z_N$ are given in \cite{Lutz:2020dfi}. 
 
Like in a computation of $G_A(q^2)$ the leading order LEC $g_A, h_A, f_S$ but also some subleading order LEC  $g_R, g^+_\chi$, but now in addition $g_\chi^-$, are needed in (\ref{res-FA}). Given the specific form of the form factor in (\ref{res-FA}) it is straight forward to match our convention for the LEC  to other choices in the literature.  
Further two-body LEC $g_S,g_V, g_T, g_R$ and $f^\pm_A,f_M$ are involved in  the loop functions, which we will specify in terms of their integrands 
\begin{eqnarray}
J^\mu_{\cdots} (\bar p, p) =  i\, \mu^{4-d}
\int \frac{d^d l}{(2\pi)^{d}} \,
K^\nu_{\cdots} (\bar p, p; l)\,\Big(  g_{\nu}^{\;\mu} - q_\nu \,q^\mu/(q^2-m_\pi^2) \Big) \,.
\label{def-Js}
\end{eqnarray}
From our previous works \cite{Lutz:2020dfi,Sauerwein:2021jxb} one can find
\begin{eqnarray}
&& K_\pi^\mu= - \frac{1}{\sqrt{2}} \,
\frac{ \gamma^{\mu} \, \gamma_5}{l^2-m_\pi^2} 
 + \frac{1}{\sqrt{2}} \,\frac{q^\mu }{3\,m_\pi^2} \,\frac{4\,M_N}{l^2-m_\pi^2}\, \gamma_5\,,
\nonumber\\
&& K^\mu_{ \pi N} =  -\frac{1}{\sqrt{2}}\,\Big( \gamma^{\mu}  + 2\,g_S\,l^\mu 
+ 2\,g_T\, i \, \sigma^{\mu\nu} \, l_{\nu} - 16\,\sqrt{3}\, g_F\, i \, \sigma^{\mu\nu} \, q_{\nu}
\nonumber\\
&& \qquad +\,\frac{1}{2}\,g_V \,
\big[ (\gamma^{\mu} \, {\bar p}\cdot l+\slashed{l}\, {\bar p}^{\mu}) + (\gamma^{\mu} \, (p-l)\cdot l+\slashed{l}\,(p-l)^{\mu})\big] \Big)\, S_N(p-l) \,\frac{\slashed{l}\,\gamma_5}{l^2-m_\pi^2}  \,,
\nonumber\\
&& K^\mu_{ N \pi} =  -\frac{1}{\sqrt{2}}\,\frac{\slashed{l}\,\gamma_5}{l^2-m_\pi^2} \, S_N(\bar{p}-l) \, \Big( \gamma^{\mu}  + 2\,g_S\,l^\mu 
- 2\,g_T\, i \, \sigma^{\mu\nu} \, l_{\nu}
\nonumber\\
&& \qquad +\,\frac{1}{2}\,g_V \,
\big[ (\gamma^{\mu} \, p\cdot l+\slashed{l}\, p^{\mu}) + (\gamma^{\mu} \, (\bar{p}-l)\cdot l+\slashed{l}\,(\bar{p}-l)^{\mu})\big]+ 16\,\sqrt{3}\, g_F\, i \, \sigma^{\mu\nu} \, q_{\nu} \Big) \,,
\nonumber\\ 
&& K^\mu_{N \pi N} = - \frac{1}{\sqrt{2}} \, \slashed{l}  \gamma_5  \, S_N(\bar{p}-l) \,\frac{ \gamma^{\mu}  \gamma_5}{l^2-m_\pi^2} \, S_N(p-l) \, \slashed{l} \gamma_5 \, ,
\nonumber\\ 
&& K^\mu_{\pi \Delta} =  -\frac{1}{\sqrt{2}}\,
 \frac{\gamma_5 \,l_\nu}{l^2-m_\pi^2}
 \Big( (f_A^- -5\,f_A^+ )\, \slashed{l} \, S_{\Delta}^{\mu\nu}(p-l)
 -(f_A^-+5\,f_A^+)\,\gamma^\mu \, l_\alpha \, S_{\Delta}^{\alpha\nu}(p-l)
\nonumber\\ 
&&\qquad -\, 4 \, f_M \, q_\alpha \, \big[\gamma^{\alpha} \, S_{\Delta}^{\mu\nu}(p-l)-\gamma^\mu \, S_{\Delta}^{\alpha\nu}(p-l)\big]\Big) \,,
 \nonumber\\
&& K^\mu_{\Delta \pi} =  \frac{1}{\sqrt{2}}\,
\Big( (f_A^- 
-5 \,f_A^+ )\, S_{\Delta}^{\nu\mu}(\bar{p}-l) \, \slashed{l}- (f_A^ -+5\,f_A^+)\,S_\Delta^{\nu\alpha}(\bar{p}-l) \, l_{\alpha} \, \gamma^\mu
\nonumber\\ 
&&\qquad
- \,4 \, f_M \, \big[S_{\Delta}^{\nu\alpha}(\bar{p}-l) \, \gamma^\mu-S_{\Delta}^{\nu\mu}(\bar{p}-l) \, \gamma^\alpha\big] \, q_\alpha
\Big) \,
\frac{l_\nu \, \gamma_5}{l^2-m_\pi^2} \,,
\nonumber\\
&&K^\mu_{N\pi \Delta} =   \frac{1}{\sqrt{2}}\,\frac{\slashed{l}\gamma_5 \,l_\tau}{l^2-m_\pi^2}\,S_N(\bar{p}-l)\,\Big(  S_{\Delta}^{\mu\tau}(p-l) + 2\,\frac{f_E}{f_S}\,\big[\gamma_{\mu}S_{\Delta}^{\alpha\tau}(p-l)-\gamma^\alpha \, S_R^{\mu\tau}(p-l)\big]\,q_\alpha\Big) \,,
\nonumber\\
&&K^\mu_{\Delta \pi N} =   \frac{1}{\sqrt{2}}\,
\Big( S_{\Delta}^{\sigma\mu}(\bar{p}-l) + 2 \, \frac{f_E}{f_S} \, \big[S_{\Delta}^{\sigma\mu}(\bar{p}-l) \, \gamma^{\alpha}-S_{\Delta}^{\sigma\alpha}(\bar{p}-l) \, \gamma^\mu\big] \, q_\alpha\Big) \, S_N(p-l)\frac{l_\sigma\,\slashed{l}\gamma_5}{l^2-m_\pi^2} \,,
\nonumber\\
&&K^\mu_{\Delta \pi \Delta} =  - \frac{1}{\sqrt{2}}\,l_{\sigma}\,S_\Delta^{\sigma\alpha}(\bar{p}-l) \,\frac{\gamma^\mu\,\gamma_5}{l^2-m_\pi^2}\, S^\Delta_{\alpha\beta}(p-l) \,l^{\beta}\,,
\label{def-Ks}
\end{eqnarray}
with the baryon propagators
\begin{eqnarray}
&&S_{N}(k)=\frac{1}{\slashed{k}-M_N }\,,
 \nonumber\\
&&  S_{\Delta}^{\mu\nu}(k)=\frac{-1}{\slashed{k}-M_{\Delta}}
\bigg(g^{\mu\nu}-\frac{\gamma^{\mu}
\gamma^{\nu}}{d-1}+\frac{(k^{\mu}\gamma^{\nu}-k^{\nu}\gamma^{\mu})}{(d-1)\,M_{\Delta}}-
\frac{(d-2)k^{\mu}k^{\nu}}{(d-1)\,M_{\Delta}^2}\bigg).
\end{eqnarray}
The main target of our current work is the derivation of the induced pseudo-scalar form factor $G_P(q^2)$ as implied by (\ref{def-Ks}) and properly expanded into its chiral moments.  We will apply our counting rules formulated in terms of on-shell hadron masses \cite{Lutz:2020dfi,Sauerwein:2021jxb}.

\clearpage
\section{Chiral expansion of the form factor}

In this section we will express the loop functions as introduced with (\ref{def-Ks}) in terms of a generalized Passarino--Veltman reduction scheme \cite{Passarino:1978jh,Lutz:2020dfi,Isken:2023xfo}. 
In a first step we apply the projection scheme (\ref{Axialcurrent}) that avoids the need to consider 
tensor-type loop integrals. Such a projection is always possible. We write
\begin{eqnarray}
&& G_A(t) = 
 g_A\,Z_N+ 4\,g^+_\chi \,m_\pi^2  + g_R \,t
\nonumber\\
&& \quad\;\, +\,\frac{g_A}{f^2}\,  \Big\{J^A_{\pi}(t) +  J_{\pi N}^A(t)+ J_{ N \pi }^A(t)\Big\}
+ \frac{g_A^3}{4\,f^2}\,J^A_{N\pi N}(t) 
 + \frac{5\,h_A\,f_S^2}{9\,f^2}\,J^A_{\Delta \pi \Delta}(t)
  \nonumber\\
&&   \quad \;\,+\, \frac{f_S}{3\,f^2}\,  \Big\{ J_{\pi \Delta}^A (t)+ J_{ \Delta \pi }^A(t)\Big\} + 
 \frac{2\,g_A\,f_S^2}{3\,f^2}\,\Big\{J^A_{N\pi \Delta }(t)+ J^A_{\Delta \pi N }(t)\Big\} + {\mathcal O} \left( Q^4\right) \,,
 \label{res-GA}
 \end{eqnarray}
and
\begin{eqnarray}
&& \frac{t-m_\pi^2}{4\,M_N^2}\,G_P(t) = 
 -g_A\,\Big(Z_N + Z_\pi+  f_\pi/f - 2 \Big)- \,m_\pi^2\,(4\,g^+_\chi + g^-_\chi)  - g_R \,(t-m_\pi^2) 
\nonumber\\
&& \quad\;\, +\,\frac{g_A}{f^2}\,  \Big\{J^P_{\pi}(t) +  J_{\pi N}^P(t)+ J_{ N \pi }^P(t)\Big\}
+ \frac{g_A^3}{4\,f^2}\,J^P_{N\pi N}(t) 
 + \frac{5\,h_A\,f_S^2}{9\,f^2}\,J^P_{\Delta \pi \Delta}(t)
  \nonumber\\
&&   \quad \;\,+\, \frac{f_S}{3\,f^2}\,  \Big\{ J_{\pi \Delta}^P (t)+ J_{ \Delta \pi }^P(t)\Big\} + 
 \frac{2\,g_A\,f_S^2}{3\,f^2}\,\Big\{J^P_{N\pi \Delta }(t)+ J^P_{\Delta \pi N }(t)\Big\} + {\mathcal O} \left( Q^4\right) \,,
 \label{res-GP}
 \end{eqnarray}
where for any loop function $J^\mu_{\cdots}(\bar p, p) \to  4\,M_N^2\,J^P_{\cdots}(t)/(t-m_\pi^2)$ and 
$J^\mu_{\cdots}(\bar p, p) \to  J^A_{\cdots}(t)$  we introduced their suitably projected forms.
It is convenient to monitor the consequences of the chiral Ward identities, which imply in particular a correlation of the pseudo-scalar and axial-vector loop functions. In the chiral limit it should hold
\begin{equation}
	\lim_{m\rightarrow 0 }\Big[G_A(t)+ \frac{t}{4\,M_N^2}\,G_P(t)\Big] =0\,
	\to \qquad \lim_{m\rightarrow0}\Big[J^A_{\ldots} + J^P_{\ldots}\Big] = 0 \,,
\label{ward-identity-PA}	
\end{equation}
which we verified by explicit computations in our renormalization scheme. 

The contributions to the form factor $G_A(t)$ and $G_P(t)$ are computed in application of a novel reduction scheme \cite{Isken:2023xfo}, in terms of scalar loop functions only. 
Recently the Passarino--Veltman scheme scheme was supplemented systematically by a set of additional basis functions that leads to expressions free of kinematical constraints and that comply with the  expectation of dimensional counting rules \cite{Isken:2023xfo}. Such an extension is required once triangle loop contributions are considered. While in our previous work 
\cite{Lutz:2020dfi} the Passarino--Veltman set was extended 
by one specific loop function in an evaluation of $G_A(t)$ only, it turned out that the more general scheme as proposed in  \cite{Isken:2023xfo} is required for $G_P(t)$. Our truncated expressions are implied by an expansion of the coefficient functions in chiral moments
\begin{eqnarray}
t\sim  m_\pi^2 \sim Q^2 \,,\qquad \delta= M_\Delta - M_N \,\Big( 1+ \frac{\Delta}{M} \Big) \sim Q^2 \,,
\end{eqnarray}
where we keep the on-shell masses unexpanded. The chiral limit value of the nucleon and isobar masses are $M$ and $M+ \Delta$ respectively. 

While we detail all axial-vector loop functions in the Appendix, the so-far unknown pseudo-scalar ones are given here 
\begin{eqnarray}
	&& \bar J^P_{\pi } (t) =  \frac{1}{3}\,\bar I_\pi + {\mathcal O}\big( Q^4\big) \,,
	\nonumber\\
	%%%
	% Beitrag Bubble N
	%%%
	&&  \bar J_{\pi N}^P (t) + \bar J_{ N \pi }^P (t)=m_\pi^2\,  \Big(1 - \frac{8}{3}\,M_N\,( g_S-2\,g_T ) \Big)\, \bar{I}_{\pi N}  + {\mathcal O}\big( Q^4\big)\,,
	\nonumber\\
	%%%
	% Beitrag Dreieck N pi N
	%%%
	&&  \bar J_{N \pi N }^P (t) = -\bar I_\pi 
	-m_\pi^2\, \bar I_{\pi N}- 4\,m_\pi^2\,M_N^2\,\Big( \bar I^{(2,0)}_{N \pi N}(t) +  \bar I^{(0,2)}_{N \pi N}(t)\Big)+ {\mathcal O}\big( Q^4\big) \,,
	\nonumber\\
	%%%
	% Beitrag Bubble Delta
	%%%
	&&  \bar J_{\pi \Delta }^P (t) + \bar J_ {\Delta  \pi }^P (t)= 
	\frac{8}{9}\,\bigg(
	25\,f_A^+\,\Big[m_\pi^2\,\alpha^P_{1 2} - 2\,\delta\,M_N\,r\,\alpha^P_{1 3}\Big]
	+f_A^-\Big[m_\pi^2\,\alpha^P_{2 2}-2\,\delta\,M_N\,r\,\alpha^P_{2 3}\Big]
	\nonumber \\
	&&\qquad \qquad \,
	-\, 8\,f_M\,r^2\,\Big[t\,\alpha^P_{3 1} - m_\pi^2\,\alpha^P_{3 2}\Big]
	\bigg)\,M_N\,\bar{I}_{\pi\Delta}  + {\mathcal O}\big( Q^4\big)	\,,
	\nonumber\\
	%%%
	% Beitrag Dreiekc N pi D + D pi N
	%%%
	&&  \bar J_{N \pi \Delta }^P (t) + \bar J_ {\Delta  \pi N }^P (t)=
	\frac{16}{9\,r}\,m_\pi^2\,\Big(\bar{I}_{\pi N} - \alpha_{42}^P\, \bar I_{\pi \Delta}\Big)
	\nonumber \\
	&&\qquad \,
	+\,\frac{2}{9}\,\bigg( r\,t\,\alpha^P_{4 1}
	+8\,\delta\,M_N\,\alpha^P_{4 3}
	+12\,\frac{f_E\,M_N\,r}{f_S}\,\Big[t\,\alpha^P_{5 1}-m_\pi^2\,\alpha^P_{5 2}\Big]
	\bigg)\,\bar{I}_{\pi\Delta}
	\nonumber \\
	&&\qquad 
		-\,\frac{8}{3}\,\frac{f_E\,M_N\,r}{f_S}\,\Big[t\,\alpha^P_{6 1} - m_\pi^2\,\alpha^P_{6 2}\Big]\,M_N^2 \,\Big(\bar{I}^{(1,0)}_{\Delta\pi N}(t) + \bar{I}^{(0,1)}_{N\pi\Delta}(t)\Big)
	\nonumber \\
	&&\qquad
	%%%
	-\,\frac{4}{9} \,\Big[ t\,\alpha^P_{7 1} - 3\,m_\pi^2\, \alpha^P_{7  2}\Big]\,M_N^2\,
	\Big( \bar{I}^{(2,0)}_{\Delta\pi N}(t)+ \bar{I}^{(0,2)}_{N\pi\Delta}(t)  \Big)
	 + {\mathcal O}\big( Q^4\big)\,,
	%%%
	\nonumber \\
	% Beitrag Delta pi Delta 
	%%%
	&&  \bar J^P_{\Delta \pi \Delta } (t)= 
	-\frac{2}{3}\,\bigg(
	2\,r\,t\,\alpha^P_{8  1} 
	+\frac{5}{9} \,m_\pi^2\,\alpha^P_{8  2}
	-\frac{10}{3}\,\delta\,M_N\,\alpha^P_{8  3}
	\bigg)\,\bar{I}_{\pi\Delta}
	\nonumber \\
	&&\qquad
	-\,\frac{4}{3}\,\Big[
	t\,\alpha^P_{9  1}+\frac{1}{3}\, m_\pi^2 \,\alpha^P_{9  2}\Big]\,M_N^2\,\Big( \bar{I}^{(2,0)}_{\Delta\pi\Delta}(t) +\bar{I}^{(0,2)}_{\Delta\pi\Delta}(t) \Big)  + {\mathcal O}\big( Q^4\big)\,,
\label{res-LoopGP}
	\end{eqnarray}
with $r=\Delta/M$ and $\alpha^{A,P}_{ab } \to 1$ at $r \to 0$. 
In Tab. \ref{tab:A} we provide specific values for such coefficients as needed in our global fit scenario.  The coefficients $\alpha^A_{ab}$ characterize the corresponding loop functions $\bar J^A_{\cdots}(t)$ as they are detailed in the Appendix. Particularly important are $\alpha_{82}^P$ and $\alpha_{82}^A$ that differ significantly from their limit value.

The set of basis loop functions are
\begin {eqnarray}
\label{def-scalar}
&& \bar I_\pi = \frac{m_\pi^2}{16\,\pi^2}\,\log \frac{m_\pi^2}{\mu^2}  \sim Q^2\,,  \qquad \qquad 
\nonumber\\
&&F_{L\pi R}(u,v)=m_\pi^2+u \, \big(M_L^2-m_\pi^2- (1-u) \, M_N^2\big) +v \, \big(M_R^2-m_\pi^2 -(1-v) \, M_N^2\big)  
\nonumber\\
&& \qquad \qquad \quad \!+\, u \, v \, \big(2\,M_N^2 -t\big)\,, 
\nonumber\\
&& \bar{I}_{\pi R}=
 \frac{\gamma_{N}^R-2}{16\,\pi^2} -
  \int_0^1  \, \frac{dv  }{16\,\pi^2}\,\log \frac{F_{L\pi R}(0,v) }{M_R^2} \sim Q\,,
\nonumber\\
&& \bar I^{(m,n)}_{L\pi R}(t)=- \frac{\gamma^{(m,n)}_{L\pi R}}{16\,\pi^2\,M^2}
+ \int_0^1\int_0^{1-u}  \, \frac{dv \, du \,u^m\,v^n}{16\,\pi^2\,F_{L\pi R}(u,v) } \sim Q^0\,,
 \label{def-scalar-loops}
\end{eqnarray}
where we introduced some subtraction terms for later convenience. We recall from \cite{Isken:2023xfo} that it suffices to consider $\bar I^{(0,n)}_{L\pi R}(t)$ and $\bar I^{(n,0)}_{L\pi R}(t)$  as additional basis functions. All remaining terms can be decomposed in terms of those without running into kinematical constraints or power-counting violating structures. 

\begin{table}[t]
	\begin{tabular}{c c @{\hspace{5em}} cc  }
			$\alpha_{12}^A / \alpha_{12}^P$ & $1.164\,/\,1.185$   & $\alpha_{51}^A\,,\alpha_{51}^P\,, \alpha_{52}^P$ & $1.239$ \\[.25em]
		$\alpha_{13}^A\,,\alpha_{13}^P$ & $1.194$   & $\alpha_{61}^A\,,\alpha_{61}^P\,,\alpha_{62}^P$ & $0.761$ \\[.25em]
		$\alpha_{22}^A\, / \, \alpha_{22}^P$ & $0.715\, / \, 0.612$    & $\alpha_{71}^A\,,\alpha_{71}^P\,,\alpha_{72}^P$ & $0.949$ \\[.25em]
		$\alpha_{23}^A \,, \alpha_{23}^P$ & $0.533 $  & 	$\alpha_{81}^A\,,\alpha_{81}^P$ & $1.554$ \\[.25em]
		$\alpha_{31}^A\,, \alpha_{31}^P\,, \alpha_{32}^P$ & $1.044$ & 	$\alpha_{82}^A \, / \, \alpha_{82}^P$ & $2.783 \,/\,3.481$  \\[.25em]
		$\alpha_{41}^A\,,\alpha_{41}^P$ & $0.932$ & $\alpha_{83}^A\,,\alpha_{83}^P$ & $1.399$ \\[.25em]
		$\alpha_{42}^A\,/\, \alpha_{42}^P$ & $1.276\,/\,1.250$ &	$\alpha_{91}^A\,,\alpha_{91}^P $ & $1.034$ \\[.25em]
		$\alpha_{43}^A\,, \alpha_{43}^P$ & $1.494$ & $ \alpha_{92}^P $&  $0.977$
	\end{tabular}
\centering
\caption{Values of the $\alpha_{ab}^{P}$ and $\alpha_{ab}^{A}$ in our global fit with $r = 0.343$.} \label{tab:A}
\end{table}

At $\gamma^R_N \to 0$  and $\gamma^{(m,n)}_{L \pi R} \to 0$ our renormalized loop functions follow the expectation of dimensional counting with $\Delta/M =r \sim Q $ as indicated in (\ref{def-scalar-loops}).  In this case all bubble and triangle contributions in (\ref{res-GP}) start at order $Q^3$. The only $Q^2$ term is implied by the pion tadpole contribution.  While such a counting is well justified for somewhat larger pion masses with $m_\pi \sim \Delta$, it looses its efficiency in the chiral domain with $m_\pi \ll \Delta$. Here one may integrate out the isobar degrees of freedom and expand in $m_\pi / \Delta$. In order to properly treat both domains in our scheme we  
use the subtraction terms
\begin{eqnarray}
&&a = r\,(2+ r)\,, \qquad \qquad \qquad \qquad \qquad  \gamma^\Delta_N = a\,\log \frac{ 1+a}{a}\,,\qquad \qquad \qquad  \gamma_N^N = 0\,, 
\nonumber\\
&&\gamma^{(0,0)}_{\Delta\pi N}=\gamma^{(0,0)}_{N\pi\Delta}=
 \frac{1}{a}\,\log (1+ a) +\log \frac{ 1+a}{a} \,,
\nonumber\\
&& \gamma^{(1,0)}_{\Delta\pi N}=\gamma^{(0,1)}_{N\pi\Delta}= \frac{1+a}{3\,a}-\frac{1}{3\,a^2}\,
\log (1+ a) - \frac{a}{3}\,\log \frac{ 1+a}{a} \,,
\nonumber\\
&& \gamma^{(0,1)}_{\Delta\pi N}=\gamma^{(1,0)}_{N\pi\Delta}= \frac{-2+a}{6\,a}+\frac{2+3\,a}{6\,a^2}\,
\log (1+ a) - \frac{a}{6}\,\log \frac{ 1+a}{a} \,,
\nonumber\\
&& \gamma^{(2,0)}_{\Delta\pi N}=\gamma^{(0,2)}_{N\pi\Delta}= \frac{-2 + a +a^2-2\, a^3}{10\,a^2}+\frac{1}{5\,a^3}\,
\log (1+ a) + \frac{a^2}{5}\,\log \frac{ 1+a}{a} \,,
\nonumber\\
&& \gamma^{(0,2)}_{\Delta\pi N}=\gamma^{(2,0)}_{N\pi\Delta}= \frac{-12 -24\,a + a^2 - 2 \,a^3}{60\,a^2}+\frac{6 + 5\, a\, (3 + 2 \,a )}{30\,a^3}\,
\log (1+ a) + \frac{a^2}{30}\,\log \frac{ 1+a}{a} \,,
\nonumber\\ 
&& \gamma^{(0,0)}_{\Delta \pi \Delta }=\log \frac{ 1+a}{a}\,, \quad 
\nonumber\\
&& \gamma^{(1,0)}_{\Delta \pi \Delta }=  \gamma^{(0,1)}_{\Delta \pi \Delta }= \frac{1}{2}\,\Big( 1-a\,\log \frac{ 1+a}{a} \Big) \,,\qquad \qquad 
 \nonumber\\
 && \gamma^{(2,0)}_{\Delta \pi \Delta }= 
\gamma^{(0,2)}_{\Delta \pi \Delta }= \frac{1}{6}\,\Big( 1-2\, a+2\,a^2\,\log \frac{ 1+a}{a} \Big)\,,
 \end{eqnarray}
which are crucial in the chiral domain, since if not included an explicit evaluation of a class of two-loop diagrams would be needed \cite{Long:2009wq}. Now 
we are in line with dimensional counting rules 
\begin{eqnarray}
&& \bar I_{\pi \Delta}(M_N^2) \sim \frac{m_\pi^2}{\Delta \,M} \sim Q^2 \,,\quad 
  \bar I^{(m,n)}_{N \pi \Delta}(t) \sim \frac{m_\pi}{\Delta \,M^2} \sim Q \,, \quad  \bar I^{(m,n)}_{\Delta \pi \Delta}(t) \sim \frac{m_\pi^2}{\Delta^2 \,M^2} \sim Q^2 \,, 
\end{eqnarray}
that arise in the chiral domain with $m_\pi/\Delta \sim Q$.

\section{LEC from Lattice QCD data}

We consider QCD lattice data on  two-flavor ensembles from ETMC \cite{Alexandrou:2008tn,Alexandrou:2010hf}, CLS \cite{Capitani:2017qpc,Capitani:2015sba}, and RQCD \cite{Bali:2014nma}  QCD lattice collaborations. While all such references provide the nucleon pion and nucleon masses together with the form factor $G_A(t)$ on their ensembles, only the ETMC provides in addition the isobar mass. This is unfortunate since the latter play a crucial role in the understanding of such form factors. Moreover, results on $G_P(t)$ are provided on  CLS and RQCD ensembles only. After a comparison of their ratios
\begin{eqnarray}
\frac{t-m_\pi^2}{4\,M_N^2} \, \frac{G_P(t)}{G_A(t)}\,,
\end{eqnarray}
it is evident that their results are largely incompatible for $G_P(t)$. This follows, since in our previous work we found that their results on $G_A(t)$ appear quite compatible. As explained in \cite{Capitani:2017qpc,Capitani:2015sba} the form factor $G_P(t)$ is suffering from sizeable and difficult-to-control excited state 
contamination at small pion masses.  
While we made an attempt to fit the results of \cite{Bali:2014nma} 
with our scheme we badly failed to recover their results for  $G_P(t)$.

In the following we discuss the results of global fits to the available data set, where we excluded all results for  $G_P(t)$ from \cite{Bali:2014nma}. We use the evolutionary fit algorithm GENEVA \cite{Geneva} with the recently implemented 
mpi consumer \cite{Wessner:2023bmi}. With this significant 
update it is possible now to call the evolutionary algorithm as a  heterogeneous mpi job. As compared to previous versions a much better scaling behaviour in the number of involved cores is observed. Our results are typically obtained on 2830 cores from reserved nodes on the Green Cube at GSI.

Our fit strategy was detailed already in our previous work \cite{Lutz:2020dfi}.  We adjust the LEC to our expressions for the nucleon and isobar masses, $M_N$ and $M_\Delta$ (with finite-volume effects included following Ref. \cite{Lutz:2014oxa}), and for the nucleon axial-vector form factor, $G_A(t)$ and $G_P(t)$ from (\ref{res-GA}, \ref{res-GP}) (without explicit finite-volume effects), to the lattice data. As explained, we use on-shell in-box meson and baryon masses in the loop expressions throughout this work. We coin such volume effects as \underline{implicit} volume effects mostly driven by the volume dependence of the in-box isobar mass.

\begin{table}[t]
\centering%\scriptsize
\renewcommand{\arraystretch}{1.2}
\begin{tabular}{|c | c |c c|c c|}\hline
group&scale & this work& lattice group
\\ \hline
ETMC&$a_{\beta=3.8}\,[\text{fm}]$&$0.1095(_{-0.0004}^{+0.0005})$&
$0.0995(7)$ \cite{Alexandrou:2008tn}
\\ \cline{2-4}
&$a_{\beta=3.9}\,[\text{fm}]$&$0.0941(_{-0.0005}^{+0.0002})$& $0.089(5)$ \cite{Alexandrou:2010hf}
\\ \cline{2-4}
&$a_{\beta=4.05}\,[\text{fm}]$&$0.0736(_{-0.0002}^{+0.0002})$&
$0.070(4)$ \cite{Alexandrou:2010hf}
\\ \cline{2-4}
&$a_{\beta=4.2}\,[\text{fm}]$&$0.0590(_{-0.0002}^{+0.0002})$&
$0.056(4)$ \cite{Alexandrou:2010hf}
\\ \hline
CLS&$a_{\beta=5.2}\,[\text{fm}]$&$0.0850(_{-0.0002}^{+0.0001})$&
0.079  \cite{Capitani:2017qpc}
\\ \cline{2-4}
&$a_{\beta=5.3}\,[\text{fm}]$&$0.0705(_{-0.0002}^{+0.0003})$ &
0.063 \cite{Capitani:2017qpc}
\\ \cline{2-4}
&$a_{\beta=5.5}\,[\text{fm}]$&$0.0525(_{-0.0001}^{+0.0001})$&
0.050 \cite{Capitani:2017qpc}
\\ \hline
RQCD&$a_{\beta=5.2}\,[\text{fm}]$&$0.0834(_{-0.0003}^{+0.0002})$& 
0.081 \cite{Bali:2014nma}
\\ \cline{2-4}
&$a_{\beta=5.29}\,[\text{fm}]$&$0.0704(_{-0.0003}^{+0.0002})$&
0.071 \cite{Bali:2014nma}
\\ \cline{2-4}
&$a_{\beta=5.4}\,[\text{fm}]$&$0.0598(_{-0.0005}^{+0.0003})$&
0.060 \cite{Bali:2014nma}
\\ \hline
\end{tabular}
\caption{Lattice scales as determined in our global fit.}
   \label{LatticeaFits}
\end{table}

The results depend on the pion mass and box size of a given  ensemble. It is important to have accurate values for the QCD lattice scales available on all considered lattice ensembles. Distinct QCD $\beta$ values are associated with distinct lattice scale parameters.  
We consider the various lattice scales as free parameters in our global fit where we isospin averaged empirical baryon masses as the scale setting condition. Such a strategy was successfully used in various global fits to lattice data  \cite{Lutz:2020dfi,Lutz:2014oxa,Lutz:2018cqo,Guo:2019nyp}. Our results for all lattice scales are shown in Table \ref{LatticeaFits}. The scales given by the ETMC and CLS collaborations differ significantly from our values. There is, however, a clear trend that the smaller the lattice scale, the closer our fitted scales get to the ones given by the lattice collaborations. The RQCD scales, on the other hand, can be reproduced quite accurately. 
Since the lattice set up of the CLS and RQCD groups coincide, one would expect 
identical lattice scales on the $\beta =5.2$ ensembles in Table \ref{LatticeaFits}. Within uncertainties this is the case for our results.

 \begin{table}
\centering%\scriptsize
\renewcommand{\arraystretch}{1.0}
\begin{tabular}{c c |c c| c c}\hline
LEC&Fit result&LEC&Fit result&LEC&Fit result
\\ \hline
$f \,[\text{MeV}]$&$83.43(_{-0.81}^{+0.30})$&
$b^*_\chi \,[\text{GeV}^{-1}]$&$-0.6805(_{-0.0020}^{+0.0085})$&
$g_S\,[\text{GeV}^{-1}]$&$0.9163(_{-0.0072}^{+0.0060})$
\\ \hline
$M \,[\text{MeV}]$&$893.79(_{-0.16}^{+0.55})$&
$d^*_\chi \,[\text{GeV}^{-1}]$&$-0.3224(_{-0.0118}^{+0.0103})$&
$g_V\,[\text{GeV}^{-2}]$&$-0.8096(_{-0.1784}^{+0.0792})$
\\ \hline
$M+\Delta \,[\text{MeV}]$&$1200.42(_{-0.39}^{+0.72})$&
$c_\chi \,[\text{GeV}^{-3}]$&$1.4627(_{-0.0452}^{+0.0256})$&
$g_T\,[\text{GeV}^{-1}]$&$1.5035(_{-0.0200}^{+0.0762})$
\\ \hline
$g_A$&$1.1449(_{-0.0049}^{+0.0019})$&
$e_\chi \,[\text{GeV}^{-3}]$&$1.2609(_{-0.0452}^{+0.0170})$&
$g_R \,[\text{GeV}^{-2}]$&$1.1233(_{-0.0318}^{+0.0118})$
\\ \hline
$f_S$&$1.5857(_{-0.0176}^{+0.0040})$&
$g^+_\chi \,[\text{GeV}^{-2}]$& $-3.9827(_{-0.0863}^{+0.0226})$ &
$h_S\,[\text{GeV}^{-1}]$&$0.7748(_{-0.0103}^{+0.0205})$
\\ \hline
$h^*_A$&$0.7893(_{-0.0229}^{+0.1123})$&
$g^-_\chi \,[\text{GeV}^{-2}]$&$-0.2752(_{-0.083}^{+0.034})$&
$h_V\,[\text{GeV}^{-2}]$&$1.9760 (_{-0.149}^{+0.152})$
\\ \hline
 $f_E\,[\text{GeV}^{-1}]$&$0.6949 (_{-0.3812}^{+0.3075})$&
$f_M\,[\text{GeV}^{-1}]$&$-0.7335(_{-0.4320}^{+0.0543})$&
$f^+_A\,[\text{GeV}^{-1}]$ &  $-0.0617(_{-0.0652}^{+0.0222}) $ \\ \hline 
 $l_3$&$0.0193 (_{-0.0003}^{+0.0003})$&
$l_4$&$-0.0151(_{-0.0011}^{+0.0003})$& $\mu^* $ [MeV] & 770
\\ \hline 
  $\zeta_N $&$0.2277 (_{-0.0475}^{+0.0207})$&
$\zeta_\Delta$&$-0.0180(_{-0.0044}^{+0.0022})$&
\\ \hline
\end{tabular}
\caption{Low-energy constants as determined in our fit. The $*$ parameters are not fitted to the lattice data. While $b_\chi$ and $d_\chi$ are adjusted to the isospin-averaged masses of the nucleon and the isobar at the physical point, the value of $h_A = 9\,g_A-6\,f_S$ is implied by its large-$N_c $ sum rule. The renormalization scale $\mu$ is set to a conventional value.}
   \label{LECFits}
\end{table}

We include in the fit ensembles with pion masses up to 500 MeV  and for lattice sizes with $m_\pi L \geq 4.0$. MeV. For the form factor we include the data points up to momentum transfer $t= -0.36$ GeV$^2$. 
The fit minimizes the least-squares differences $\chi^2$ of our expressions with respect to the lattice data points. In this $\chi^2$ determination, all available lattice points that meet our requirements  contribute with equal weight. The $\chi^2$ per lattice point reached is $\chi^2_{\rm min}/N_{\rm data}=0.799$. With $124$ used lattice data points and 32 degrees of freedom ( 22 LEC and 10 lattice scales), we reach for the total $\chi^2$ per degree of freedom
\begin{align}
\chi^2_{\rm min}/N_{\rm df} =99.12/(124-32) = 1.077 \, ,  
\end{align}
which signals a fair description of the available lattice data. 
As compared to \cite{Lutz:2020dfi} the improvement of the QCD lattice data reproduction is a consequence of using an updated  evaluation of the loop functions but also by considering a large set of data, that includes the form factor $G_A(t)$ \underline{and} $G_P(t)$. We search for local minima of our chisquare function where we are interested only in those minima, which have naturally sized LEC. This is easily possible within our  evolutionary algorithm. 

In Tab. \ref{LECFits} we collect the values for the LEC from our global fit. 
We give asymmetric one-sigma error bars. They are based on a one standard deviation ($\sigma$) change for the value of $\chi^2_{\rm min}$ ({\it i.e.} an increase by 1). We determined the region for the LEC meeting this range, from which follow the errors for the LEC. While our results are in qualitative agreement with previous studies there are important differences. For instance it shows an expected value for $\bar{l}_3$ \cite{Bernard:2007zu,Aoki:2019cca}. 
Also for  our values for $f$, $M$, $M+\Delta$, and $g_A$ are within range of \cite{Aoki:2019cca}. On the other hand
we find values for $g_S$ and $g_V$ that disagree significantly from previous SU(2) works like Refs. \cite{Bernard:2007zu,Gasparyan:2010xz}. However, in most papers the constants $g_S$ and $g_V$ are determined in a theory without isobars. 

A comparison of our $l_3$ and $l_4$ in Tab.  \ref{LECFits} with values listed in \cite{FlavourLatticeAveragingGroupFLAG:2021npn} reveals a striking tension. While our $l_3$ translates into an $\bar l_3$ value of about $\sim 3.07$ our value for $l_4$ has a sign opposite to typical values from  \cite{FlavourLatticeAveragingGroupFLAG:2021npn}. We scrutinized our results for $l_4$ by performing further fits  by selecting ensembles with $m_\pi < 450$ MeV but also $m_\pi < 550$ MeV. In both cases the fit quality is changed slightly only. While the LEC change somewhat outside the tiny 1-sigma error band none of the striking features of our best fit scenario in Tab. \ref{LECFits}  with $m_\pi < 500$ MeV is altered. In particular our $l_4$ remains negative always.

It is interesting to observe that in our current fit the sign of $h_A$ changed as compared to \cite{Lutz:2020dfi}. A negative sign was also obtained previously in \cite{Yao:2016vbz} from a study of loop corrections in pion-nucleon scattering. While at leading order in a  $1/N_c$ expansion one may favour a large and positive $h_A \sim 9\,g_A/5$ value, at subleading order its sign is not fixed with $h_A =9\, g_A - 6\,f_S$. Depending on the values of $g_A> 0 $ and $f_S > 0$ it may turn positive or negative. In the axial-vector form factor the contribution $\sim  h_A\,f_S^2\,J_{\pi  \Delta \pi}(t)$ probes the sign of $h_A$. A possible more direct 
strategy to determine such a phase was suggested recently in \cite{Bertilsson:2023htb}. The fact that we find a strong sensitivity of $h_A$ on the detailed form of how to incorporate the isobar degrees of freedom, we may speculate that the specifics of loop corrections in pion-nucleon scattering may be subject to similar effects. Here the novel development \cite{Isken:2023xfo} 
may turn out instrumental.

 \begin{table}
\centering%\scriptsize
\renewcommand{\arraystretch}{1.0}
\begin{tabular}{c |c }\hline
Observable&Fit results
\\ \hline
$G_A(0)$&$1.2284(_{-0.0059}^{+0.0021})$
\\ \hline
$\langle r_A^2\rangle\,[\text{fm}^2]$&$0.20137(_{-0.0035}^{+0.0032})$
\\ \hline
$g_P $&$8.2521(_{-0.039}^{+0.039})$
\\ \hline \hline 
$\sigma_{\pi N}\,[\text{MeV}]$&$42.22(_{-0.05}^{+0.02})$
\\ \hline
$\sigma_{ \pi \Delta}\,[\text{MeV}]$&$35.27(_{-0.06}^{+0.01})$
\\ \hline
$f_{ \pi}\,[\text{MeV}]$&$84.96(_{-0.82}^{+0.29})$
\end{tabular}
\caption{Observables as determined in our fit.}
   \label{ObservablesFits}
\end{table}

In Tab. \ref{ObservablesFits} we list additional observable quantities as they are implied by our set of LEC. Most interesting is our prediction for the axial-vector coupling constant, which is significantly below its empirical value $G_A(0)=1.2732(23)$ \cite{Tanabashi:2018oca}. We interpret this discrepancy as the effect from the neglected strange-quark mass effects in our approach. 

The axial radius may be compared with its empirical value  $ \langle r_A^2\rangle = 0.46(24) $ fm$^2$ \cite{Meyer:2016oeg,Hill:2017wgb}, where we find  our value to be roughly consistent with its empirical expectation. Previous lattice values show quite some spread \cite{Yao:2017fym,Alexandrou:2017hac,Capitani:2017qpc,Green:2017ke} but tend to prefer also a small radius with for instance 
$ \langle r_A^2\rangle = 0.213(6)(13)(3)  $ fm$^2$ from  \cite{Green:2017ke}. Similarly the empirical value for $g_P = 10.6(2.7)$ from \cite{Gorringe:2002xx}
\begin{eqnarray}
g_P = \frac{m_\mu}{2\,M_N}\,G_P(-0.877 \,m_\mu^2) \,,
\end{eqnarray}
is surprisingly close to our range in Tab. \ref{ObservablesFits}. So one may expect that the effect of 
strange quarks is less visible here. 

Most striking we find our value for the pion-nucleon sigma term 
\begin{align}
 \sigma_{ \pi N}=m\, \frac{\partial}{\partial \,m}\, M_N \,,
\end{align}
and the pion decay constant  
which both show a significant conflict  
with the empirical value $\sigma_{\pi N} = 58(5)$ MeV from Ref. \cite{Hoferichter:2015dsa,Siemens:2016jwj,RuizdeElvira:2017stg} and the PDG value $f_\pi =  (92.21 \pm 0.14) $ MeV \cite{ParticleDataGroup:2022pth}. 
Our sigma term also differs significantly from our previous value obtained \cite{Lutz:2020dfi}, which we take as strong hint that details how to incorporate the isobar degrees of freedom are crucial for such observables. Our current value is quite consistent with previous analysis \cite{Procura:2006bj,Alvarez-Ruso:2013fza} that obtained $\sigma_{\pi N}$ = 41(5)(4) MeV based on a flavor-SU(2) extrapolation of an older set of lattice data for the nucleon mass \cite{Bali:2012qs,Alexandrou:2010hf,Engel:2010my,Capitani:2012gj}.  While some recent fits of flavour SU(3) lattice QCD data in \cite{Lutz:2023xpi,Gupta:2021ahb} appear compatible with the empirical value, 
there is a large spread in values in the literature based on different assumptions and data sets (see e.g.  \cite{Alexandrou:2019brg,Bali:2016lvx,Lutz:2018cqo,Guo:2019nyp} ).  

This raises the question on the role of strange quarks in the sigma term but also in the pion decay constant. The reason for our discovery in the  decay constants stems to a large extent in the sign of $l_4$. In contrast to previous works based on flavour SU(2) ensembles as listed in \cite{FlavourLatticeAveragingGroupFLAG:2021npn}, in our work we determined $l_4$ from the pseudoscalar induced form factor of the nucleon for the first time.  Setting the lattice scale for SU(2) ensembles by the empirical decay constant, a rather popular scheme, may masks the potential discovery of such an effect. Lattice studies with a scale set to the nucleon mass would be much preferred in this case.

\clearpage

\section{Summary and outlook}

In our work we presented a chiral extrapolation study of lattice QCD data based on flavour SU(2) ensembles of CLS, RQCD and ETMC. Our emphasis is the role of the isobar in a computation of the axial-vector form factor of the nucleon. Here the evaluation of one-loop effects involving the isobar was worked out in a novel chiral framework, that permits a systematic subtraction of power-counting terms, but still use on-shell hadron masses.   For the first time we achieved a simultaneous reproduction on QCD lattice data for both components of that form factor on ensembles with pion masses up to 500 MeV.

Our results indicate the crucial importance of strange quark effects on the axial-vector form factor. Lattice data in the absence of active strange degrees of freedom do not seem to be able to reproduce the empirical form factors at the physical point. The axial-vector coupling constant of the nucleon is underestimated, but also the pion-nucleon sigma term turns out to be much below its empirical value. While our current fit to the available data set is already excellent it can not be ruled out that our conclusions are affected upon a full consideration of finite-box effects or 
a further improved data set.

It is an open challenge to generate and analyse further Lattice QCD data at fixed strange quark masses, which can be extrapolated faithfully to the physical pion masses and volumes. Only then 
a significant comparison with empirical results is convincing. 
Given this challenge it is of utmost importance to generate high statistic data on the isobar masses for such systems on ensembles with varying volumes and pion masses.  On our side we plan to provide a full computation of finite box effects for such form factors. We expect that with those it may be possible to also consider further discretization effects by the spurion field approach. 

\section{Acknowledgments}
We thank James Hudspith and Daniel Mohler for useful discussions on QCD lattice aspects. M.F.M. Lutz appreciates the support of Denis Bertini by  
providing a spack and slurm free container for the Green Cube at GSI that can be used for heterogeneous mpi jobs on reserved nodes. Jonas Wessner and Kilian Schwarz were instrumental to show how to use the mpi consumer of GENEVA at GSI.  
\clearpage 

\appendix
\section{Loop functions for $G_A$}
We confirm the form of the axial-vector loop functions as derived in \cite{Lutz:2020dfi}. A revision of such results is required as to arrive at results that comply with (\ref{ward-identity-PA}) and 
 (\ref{res-LoopGP}). The following form is found
\begin{eqnarray}
	&& \bar J^A_{\pi } (t) =  - \bar I_\pi + {\mathcal O}\big( Q^4\big) \,,
	\nonumber\\
	%%%
	% Beitrag Bubble N
	%%%
	&&  \bar J^A_{\pi N} (t) + \bar J_{ N \pi }^A (t)= 2\, m_\pi^2\,  \Big(-1 + \frac{4}{3}\,M_N\,( g_S-2\,g_T ) \Big)\, \bar{I}_{\pi N}  + {\mathcal O}\big( Q^4\big)\,,
	\nonumber\\
	%%%
	% Beitrag Dreieck N pi N
	%%%
	&&  \bar J^A_{N \pi N } (t) = \bar I_\pi + m_\pi^2\,\bar{I}_{\pi N}+ {\mathcal O}\big( Q^4\big) \,,
	\nonumber\\
	%%%
	% Beitrag Bubble Delta
	%%%
	&&  \bar J^A_{\pi \Delta } (t) + \bar J_ {\Delta  \pi }^A (t)= 
	\frac{8}{9}\bigg(
	25\,f_A^+\Big[-m_\pi^2\,\alpha^A_{1 2} + 2\,\delta\,M_N\,r\,\alpha^A_{1 3}\Big]
		\nonumber\\
	&& \qquad \qquad\,
	+ \,f_A^-\Big[-m_\pi^2\,\alpha^A_{2 2} + 2\,\delta\,M_N\,r\,\alpha^A_{2 3}\Big]
	+8\,f_M\,r^2\,t\,\alpha^A_{3 1}
	\bigg)\,M_N\,\bar{I}_{\pi\Delta}
	+\mathcal{O}\big(Q^4\big)\,,
	\nonumber\\
	%%%
	% Beitrag N pi Delta + Delta pi N
	%%%
	&&  \bar J_{N \pi \Delta }^A (t) + \bar J_ {\Delta  \pi N }^A (t)=
	-\frac{16}{9\,r}\,m_\pi^2\,\Big(\bar{I}_{\pi N} - \alpha_{42}^A\, \bar I_{\pi \Delta}\Big)
	\nonumber \\
	&&\qquad \,
	-\,\frac{2}{9}\,\Big(r\,t\,\alpha^A_{4 1}
	+8\,\delta\,M_N\,\alpha^A_{4 3}
	+12\,\frac{f_E\,M_N\,r}{f_S}\,t\,\alpha^A_{5 1}\Big)\,\bar{I}_{\pi\Delta}
	\nonumber \\
	&&\qquad 
		+\,\frac{8}{3}\,\frac{f_E\,M_N\,r}{f_S}\,t\,\alpha^A_{6 1}\,M_N^2 \,\Big(\bar{I}^{(1,0)}_{\Delta\pi N}(t) + \bar{I}^{(0,1)}_{N\pi\Delta}(t)\Big)
\nonumber \\
	&&\qquad		
		+\,\frac{4}{9} \, t\,\alpha^A_{7 1}\,M_N^2\,
	\Big( \bar{I}^{(2,0)}_{\Delta\pi N}(t) +  \bar{I}^{(0,2)}_{N\pi\Delta} (t)\Big)
	 + {\mathcal O}\big( Q^4\big)\,,
	\nonumber \\
	%%%
	% Beitrag Delta pi Delta 
	%%%
	&&  \bar J^A_{\Delta \pi \Delta } (t)= 
	\frac{2}{3}\,\bigg(
	2\,r\,t\,\alpha^A_{8  1} 
	+\frac{5}{9} \,m_\pi^2\,\alpha^A_{8  2}
	-\frac{10}{3}\,\delta\,M_N\,\alpha^A_{8  3}
	\bigg)\,\bar{I}_{\pi\Delta}
	\nonumber \\
	&&\qquad
	+\,\frac{4}{3}\,
	t\,\alpha^A_{9  1}\,M_N^2\,\Big( \bar{I}^{(2,0)}_{\Delta\pi\Delta}(t) +\bar{I}^{(0,2)}_{\Delta\pi\Delta}(t) \Big)  + {\mathcal O}\big( Q^4\big)\,,
\end{eqnarray}
together with the rational functions $\alpha^A_{\ldots}$ and $\alpha^P_{\ldots}$ which take the form
\begin{eqnarray}
	&&\alpha^A_{1  2} = \frac{(2+r)^2 (20+24 \,r+19 \,r^2+3 \,r^3)}{80 \,(1+r)^2}\,,
	\nonumber \\  
	%%%%
	&&\alpha^A_{1  3} = \frac{(2+r)^3 (20+36 \,r+29 \,r^2+5 \,r^3)}{160 \,(1+r)^3}\,,
	\nonumber \\ 
	%%%%
	&&\alpha^A_{2  2} = \frac{(2+r)^2 (4-r^2-3 \,r^3)}{16\,(1+r)^2}\,,
	\nonumber \\ 
	%%%%
	&&\alpha^A_{2  3} = \frac{(2+r)^3 (4-5 \,r^2-5 \,r^3)}{32 \,(1+r)^3}\,,
	\nonumber \\ 
	%%%%
	&&\alpha^A_{3  1} = \frac{(2+r)^4}{16\,(1+r)^2}\,,
	\nonumber \\ 
	%%%%
	\nonumber\\
&& \alpha^A_{4 1} =  \frac{(2+r)^2 (1+ r-\,r^2)}{4\,(1+r)^2}\,,
\nonumber \\
&& \alpha^A_{4 2} =  \frac{(2+r)^3 (4+4 \,r+3 \,r^2)}{32\,(1+r)^2}\,,
\nonumber \\
&& \alpha^A_{4 3} =  \frac{(2+r)^4 (1+2 \,r+2 \,r^2)}{16\,(1+r)^3}\,,
\nonumber \\
&& \alpha^A_{5 1} =  \frac{(2+r)^2 (3+4 \,r+4 \,r^2+ r^3)}{12\,(1+r)^2)}\,,
\nonumber \\
&& \alpha^A_{6 1} =  \frac{(2+r)^2}{4\,(1+r)^2}\,,
\nonumber \\
&& \alpha^A_{7 1} =  \frac{(2+r) (1+ r+r^2)}{2\,(1+r)^2}\,,
\nonumber \\
&& \alpha^A_{8 1} =  \frac{(2+r)^3 (18+60 \,r+29 \,r^2+4 \,r^3)}{144\,(1+r)^3}\,,
\nonumber \\
&& \alpha^A_{8 2} =  \frac{(2+r) (20+212 \,r+398 \,r^2+325 \,r^3+130 \,r^4+19 \,r^5)}{40\,(1+r)^4}\,,
\nonumber \\
&& \alpha^A_{8 3} =  \frac{(2+r)^2 (30+130 \,r+319 \,r^2+418 \,r^3+296 \,r^4+107 \,r^5+14 \,r^6)}{120\,(1+r)^5}\,,
\nonumber \\
&& \alpha^A_{9 1} =  \frac{(2+r)^2 (6+6\,r+r^2)}{24\,(1+r)^2}\,.
	\nonumber \\   \nonumber\\
	&&\alpha^P_{1  2} = \frac{(2+r)^2 (40+48 \,r+42 \,r^2+18 \,r^3+9 r^4+2 r^5)}{160\,(1+r)^2}\,,
	\nonumber \\  
	%%%%
	&&\alpha^P_{1  3} = \frac{(2+r)^3 (20+36 \,r+29 \,r^2+5 \,r^3)}{ 160\,(1+r)^3}\,,
	\nonumber \\ 
	%%%%
	&&\alpha^P_{2  2} = \frac{(2+r)^2 (8-6 \,r^2-18 \,r^3-9 \,r^4-2 r^5)}{32 \,(1+r)^2}\,,
	\nonumber \\ 
	%%%%
	&&\alpha^P_{2  3} = \frac{(2+r)^3 (4-5 \,r^2-5 \,r^3)}{32\,(1+r)^3}\,,
	\nonumber \\ 
	%%%%
	&&\alpha^P_{3  1} = \frac{(2+r)^4}{16\,(1+r)^2}\,,
	\qquad \qquad \qquad \qquad \qquad \quad 
	\alpha^P_{3  2} =\alpha^P_{3  1} \,,
	\nonumber \\ 
	%%%%
&& \alpha^P_{4 1} =  \frac{(2+r)^2 (1+ r-\,r^2)}{4\,(1+r)^2}\,,
\nonumber \\
&& \alpha^P_{4 2} =  \frac{(2+r)^2 (8+12 \,r+7 \,r^2+5 \,r^3+ r^5)}{32\,(1+r)^2}\,,
\nonumber \\
&& \alpha^P_{4 3} =  \frac{(2+r)^4 (1+2 \,r+2 \,r^2)}{16\,(1+ r)^3}\,,
\nonumber \\
&& \alpha^P_{5 1} =  \frac{(2+r)^2 (3+4 \,r+4 \,r^2+ r^3)}{12\,(1+r)^2}\,, \qquad \qquad  \alpha^P_{5 2} =  \alpha^P_{5 1}\,,
\nonumber \\
&& \alpha^P_{6 1} =  \frac{(2+r)^2}{4\,(1+r)^2}\,, \qquad \qquad \qquad 
\qquad \qquad \qquad 
\alpha^P_{6 2} =  \alpha^P_{6 1}\,,
\nonumber \\
&& \alpha^P_{7 1} =  \frac{(2+r) (1+ r+r^2)}{2\,(1+r)^2}\,,
\qquad \qquad \qquad \qquad \alpha^P_{7 2} =  \alpha^P_{7 1}\,,
\nonumber \\
&& \alpha^P_{8 1} =  \frac{(2+r)^3 (18+60 \,r+29 \,r^2+4 \,r^3)}{144\,(1+r)^3}\,,
\nonumber \\
&& \alpha^P_{8 2} =  \frac{(2+r) (20+236 \,r+518 \,r^2+603 \,r^3+442 \,r^4+187 \,r^5+40 \,r^6+3 \,r^7)}{40\,(1+r)^4}\,,
\nonumber \\
&& \alpha^P_{8 3} =  \frac{(2+r)^2 (30+130 \,r+319 \,r^2+418 \,r^3+296 \,r^4+107 \,r^5+14 \,r^6)}{120\,(1+r)^5}\,,
\nonumber \\
&& \alpha^P_{9 1} =  \frac{(2+r)^2 (6+6 \,r+r^2)}{24\,(1+r)^2}\,,
\nonumber \\
&& \alpha^P_{9 2} =  \frac{3(2+r)^2 (2+2 \,r-\,r^2)}{24\,(1+r)^2}\,.
	\end{eqnarray}
The expectation (\ref{ward-identity-PA}) implies specific relations amongst the coefficients
\begin{eqnarray}
&& \alpha^A_{n1} = \alpha^P_{n1} \qquad {\rm for}\,\qquad  n= 1, \cdots, 9\,,  \qquad {\rm and} \qquad  \alpha^P_{32} =\alpha^P_{31} \,,\quad  \alpha^P_{52} =\alpha^P_{51} \,,\quad 
 \alpha^P_{62} =\alpha^P_{61} \,,
\nonumber\\
&& \alpha^A_{n3} = \alpha^P_{n3}  \qquad {\rm for}\,\qquad  n= 1, \cdots, 9\,,
\label{alpha-Ward-Identify}
\end{eqnarray}
which is indeed verified by our explicit computations.  Note that the identities in the second line (\ref{alpha-Ward-Identify}) are accidental and not related to the chiral Ward identities.

\newpage
\bibliography{literature_revised} 

\end{document}